\documentclass{article}
\usepackage{spconf,amsmath,epsfig}
\usepackage{amsfonts}
\usepackage{svg}
\usepackage{url}

\title{DeepSUM++: Non-local Deep neural network for Super-resolution of Unregistered Multitemporal images}

\name{Andrea Bordone Molini, Diego Valsesia, Giulia Fracastoro, Enrico Magli\thanks{This research has been funded by the Smart-Data@PoliTO center
for Big Data and Machine Learning technologies.}}
\address{Politecnico di Torino, Italy}
\begin{document}
%\ninept
%
\maketitle
\begin{abstract}
Deep learning methods for super-resolution of a remote sensing scene from multiple unregistered low-resolution images have recently gained attention thanks to a challenge proposed by the European Space Agency. This paper presents an evolution of the winner of the challenge, showing how incorporating non-local information in a convolutional neural network allows to exploit self-similar patterns that provide enhanced regularization of the super-resolution problem. Experiments on the dataset of the challenge show improved performance over the state-of-the-art, which does not exploit non-local information.
\end{abstract}
\begin{keywords}
Super-resolution, CNN, Non-local
\end{keywords}
\vspace{-6pt}
\section{Introduction}
\label{sec:intro}
\vspace{-6pt}

The task of generating a high resolution (HR) image from one or more low resolution (LR) images is named super-resolution (SR). Despite the continuous development of ever more advanced optical devices, for some imaging applications acquiring HR images can be cumbersome or impossible due to theoretical and practical limitations. Super-resolution methods are thus required to improve image resolution beyond the sensor capability.
The approaches to image super-resolution can be framed into two main categories: single-image SR (SISR) and multi-image SR (MISR).
SISR exploits spatial correlation in a single image to recover the HR version. Certain scenarios provide multiple LR versions of the same scene, which can be combined by means of MISR techniques. In MISR, the reconstruction of high spatial-frequency details takes full advantage of the complementary information coming from different observations of the same scene.
MISR was first explored by Tsai and Huang \cite{tsaiHuang1984} who %used multiple under-sampled images with sub-pixel displacements to increase the resolution of Landsat TM images in frequency-domain.
increased the resolution of Landsat TM images in frequency-domain.
Many other MISR techniques were proposed over the years \cite{935034} mainly in spatial domain to overcome the drawbacks affecting the frequency-domain. Typical spatial-domain methods include non-uniform interpolation \cite{1176931}, iterative back-projection (IBP) \cite{IRANI1991231}, projection onto convex sets (POCS) \cite{Stark:89}, regularized methods \cite{1331445}, and sparse coding \cite{KATO201564}.
Some of these works  have  been  proposed for remote  sensing  applications. 
To improve the effectiveness of MISR methods, non-local techniques have also been introduced by Protter et al. in \cite{4694003,4914799}, based on the non-local means filter \cite{1467423}. The idea of these methods is to exploit non-local structural similarity across spatially distant patches within an image.
%Other non local MISR techniques (Video super resolution based on non-local regularization and reliable motion estimation) focus on improving regularization of the HR image reconstruction directly using non local priors. 

%In the past decades, many kinds of regularizers have been proposed to preserve edge information while removing image noise, such as Tikhonov regularizer \cite{Hardie98,913592}, Markov random field regularizer \cite{6096366}, total variation (TV) \cite{661187,Marquina2008,Zhang2014} and  bilateral total variation (BTV) \cite{1331445}. The problem is that all these are based on the general statistics of natural images, inducing a stair-casing effect in the final HR reconstruction. The nonlocal-based model can make use of more information taking into account the peculiar characteristics of the image to be recovered.

Recently, deep learning techniques have spread in various fields of research, due to their ability of  extracting high-level features from images. For SISR, deep learning works \cite{DnCnnZhang,Zhang2018ResidualDN} %liu2018non,10.1007/978-3-319-10593-2_13,kim2016deep,kim2015deep_rec,Shi2016RealTimeSI,Lim2017EnhancedDR,Zhang2018ResidualDN}%
achieved state-of-the-art results both on traditional natural images and remote sensing imagery \cite{8600724}.
Instead, just a few works propose deep learning solutions addressing MISR for remote sensing imagery \cite{SuperDeep_MISR,deepsum}, which is particularly challenging due to the highly detailed images and complex statistics. Addressing MISR from multiple low-resolution acquisitions of the PROBA-V satellite over the span of a month was the goal of a recent challenge issued  by the European Space Agency \cite{web:kelvins}.
The winning approach was DeepSUM \cite{deepsum}: a convolutional neural network (CNN) architecture that can handle registration and fusion of multiple unregistered images from the same scene in way that is robust to content variations. Its main novelty is that it can learn to concurrently solve the reconstruction and registration tasks to obtain a single HR image.

Exploiting non-local information in CNN architectures is currently a hot research topic. Some works in the denoising literature \cite{liu2018non,valsesia2019deep} attempt various approaches to define CNNs able to combine both spatially-neighboring as well as distant pixels. The graph convolution in \cite{valsesia2019deep} adopts a non-local convolution operator where a similarity graph is constructed connecting pixels whose feature vectors are close to each other. 

In this paper, we introduce graph-convolutional layers in the DeepSUM architecture, in order to improve its learning capability and generate non-local feature maps in the hidden layers to solve a MISR problem on remote sensing imagery. The resulting DeepSUM++ architecture shows that the performance of DeepSUM is greatly enhanced by the non-local operations, improving upon the state of the art. 

\begin{figure*}[t]
\centering
\includegraphics[width=0.9\textwidth]{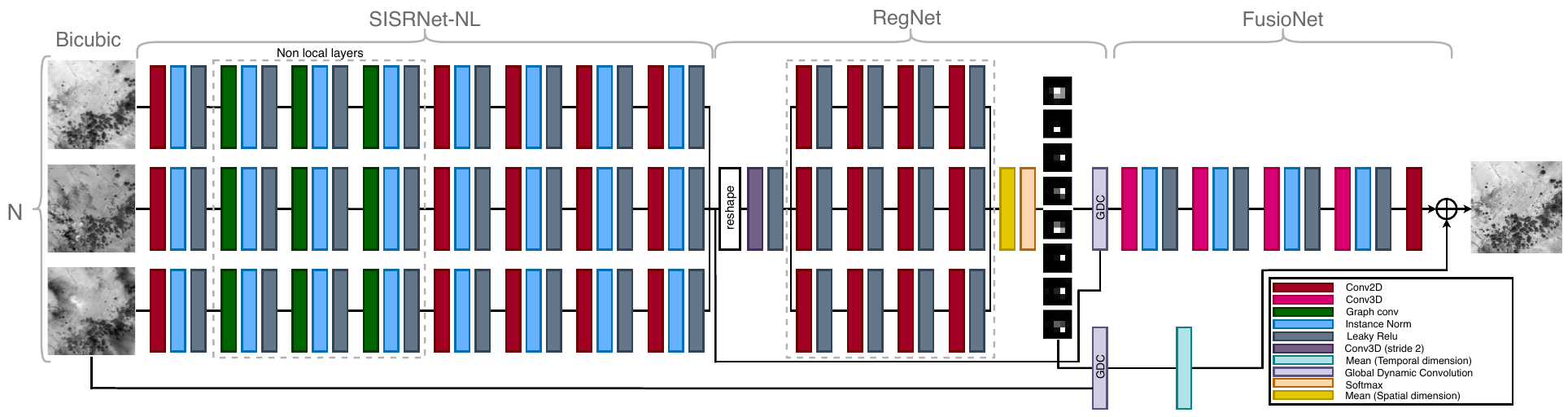}
\vspace{-0.4cm}
\caption{DeepSUM++ architecture. Graph-convolutional layers are used in SISRNet-NL.}
\vspace{-0.3cm}
\label{fig:Architecture}
\end{figure*}

\vspace{-8pt}
\section{Proposed method}
\label{sec:method}
\vspace{-8pt}

The proposed method, called DeepSUM++, reconstructs a high-resolution image $I^\text{HR}$ given a set of $N$ LR images  $I_{[0,N-1]}^\text{LR}$ representing the same scene.
$I_{[0,N-1]}^\text{LR}$ and $I^\text{HR}$ are represented as real-valued tensors with shape $N \times H \times W \times C$ and $1 \times rH \times rW\, \times C$ respectively, where $H$ and $W$ are the height and the width of the input LR frames, %$N$ is the number of LR images referred to a specific scene stacked along a third “temporal” dimension, 
$C$ is the number of channels and $r$ is the scale factor. 

DeepSUM++ solves a MISR problem in a setting with multitemporal LR image acquisitions. This is characterized by variations in scene content over multiple acquisitions due to weather or human activities. Moreover, the absolute image brightness may vary among LR images as well as between the reference HR image and the LR set. Finally, the LR images are not registered with each other, and we assume that the geometric disparity is only translational.  

DeepSUM addressed this problem with a CNN composed of three main building blocks: feature extraction (SISRNet), registration in feature space (RegNet) and fusion to obtain a single HR reconstruction (FusionNet). All blocks are based on classical 2D or 3D convolutions, so only local information is exploited to obtain the final HR reconstruction. DeepSUM is optimized in an end-to-end fashion allowing the registration and fusion task to leverage the feature learning capabilities of SISRNet that is in fact a crucial component.

DeepSUM++ builds upon the DeepSUM architecture and introduces non-local operations in the SISRNet block, which allows SISRNet to compute more powerful high-dimensional image representations that considerably improve the quality of the subsequent tasks by relying on more informative features. An overview of the network is shown in Fig. \ref{fig:Architecture}. %turning SISRNet into SISRNet-NL

The first block, called SISRNet-NL (non local), is a simple SISR network without the output projection to a single channel, where $N$ bicubically interpolated LR (ILR) images are processed independently. The weights are shared along the temporal dimension, i.e., all the $N$ ILR images go through the same operators. Overall, it acts as a feature extractor exploiting both local and non-local spatial correlations to improve upon the initial bicubic interpolation. In order to exploit non-local spatial correlation, traditional convolution is augmented with a graph convolution operator, which adds to the receptive field of a pixel a predefined number of non-local pixels whose feature vectors are close to the feature vector of the current pixel.

More formally, the graph-convolutional layer takes as input the feature vectors $\textbf{H}^l \in \mathbb{R}^{F^l \times rHrW} $, i.e., the high-dimensional representations of the image pixels at layer $l$, and the adjacency matrix of a graph connecting image pixels. The graph structure is constructed as a $K$-nearest neighbor graph where each pixel is connected to the $K$ pixels whose feature vectors are closest in terms of Euclidean distance, within a search window. The 8 local neighbors are excluded from this search as they already contribute to the local convolution.
The graph-convolutional layer is composed of two components both taking as input the feature vectors  $\textbf{H}^l $. A classic 2D convolution aggregates the local neighbors through $3 \times 3$ filters and an edge-conditioned convolution (ECC) \cite{simonovsky2017dynamic,valsesia2019deep} aggregates the feature vectors of the non-local pixels. For each pixel $i$, the ECC computes the output feature vector $\mathbf{H}^{l+1,\text{NL}}_i\in\mathbb{R}^{F^{l+1}}$ as follows:
\begin{align*}
\mathbf{H}^{l+1,\text{NL}}_i \hspace{-2pt} &= \hspace{-6pt} \sum_{j\in\mathcal{N}^l_i} \hspace{-4pt} \gamma^{l,j\xrightarrow{}i} \frac{\mathcal{F}^l_{w^l} (\mathbf{d}^{l,j\xrightarrow{}i}) \mathbf{H}^l_j}{|{\mathcal{N}^l_i}|} \hspace{-2pt}
= \hspace{-4pt} \sum_{j\in\mathcal{N}^l_i} \hspace{-4pt} \gamma^{l,j\xrightarrow{}i} \frac{\Theta^{l,j\xrightarrow{}i} \mathbf{H}^l_j}{|{\mathcal{N}^l_i}|}
\end{align*}
where $\mathcal{F}^l_{w^l}:  \mathbb{R}^{F^{l}} \xrightarrow{} \mathbb{R}^{F^{l+1} \times F^l}$ is a fully-connected network, parameterized by $w^l$, that takes as input the differences between feature vectors $\mathbf{d}^{l,j\xrightarrow{}i}=\mathbf{H}^l_j-\mathbf{H}^l_i$ and outputs a weight matrix, and $\mathcal{N}^l_i$ is the set of non-local neighbors of pixel $i$. $\gamma^{l,j\xrightarrow{}i}$ is a non-learnable scalar edge-attention term to underweight the edges between pixels with distant feature vectors for training stabilization. This term is computed as:
\vspace{-4pt}
\begin{align*}
\gamma^{l,j\xrightarrow{}i} = \exp (-\Vert{\mathbf{d}^{l,j\xrightarrow{}i}}\Vert^2_2 / \delta)\\[-20pt]
\end{align*}
where $\delta$ is a hyperparameter.

Finally, the local and the non-local contributions are averaged to estimate the output feature vector:
\begin{align*}
\mathbf{H}^{l+1}_i =\frac{\mathbf{H}^{l+1,\text{NL}}_i + \mathbf{H}^{l+1,\text{L}}_i}{2} + \mathbf{b}^l 
\end{align*}
where $\mathbf{H}^{l+1,\text{L}}_i$ is the 2D convolution output and $b^l$ is a bias. We refer the reader to \cite{valsesia2019deep} for more details on the advantages of this definition of graph convolution.
Depending on the desired computational complexity one may want to use graph convolution for all the layers in SISRNet or just for a subset.

The second network block, called RegNet, dynamically estimates a set of 2D filters using the $N$ higher dimensional image representations produced by the SISRNet-NL block to register them to each other. Handling registration within an end-to-end trainable network enables the generation of adaptive filters for disparity compensation.

The third block, called FusionNet, merges the registered image representations in the feature space in a ``slow'' fashion, i.e., by exploiting a sequence of 3D convolutional operations with small kernels. This block allows the network to learn the best space to decouple image features that are relevant to the fusion from irrelevant features and to construct the best function to exploit spatio-temporal correlations. The output of this block is a single super-resolved image.

Finally, the proposed architecture employs a global input-output residual connection. The network estimates only the high frequency details necessary to correct the bicubically-upsampled input. This is an established technique for image restoration problems \cite{DnCnnZhang}, including SISR. However, with respect to SISR, our proposed network is a many-to-one mapping, so the residual is actually added element-wise to a basic merge of the registered ILR input images $I_{[0,N-1]}^\text{IRLR}$ in the form of their average. Notice that registration of the input images $I_{[0,N-1]}^\text{ILR}$ is performed before averaging by means of the same filters produced by RegNet as illustrated in Fig. \ref{fig:Architecture}. Hence, the output is computed as follows:
\begin{align*}
   I^\text{SR} &= \frac{1}{N}\sum_{i\in[0,N-1]}{I_{i}^\text{IRLR}} + R.
\end{align*}
being $R$ the residual estimated by the CNN.

\vspace{-8pt}
\subsection{Loss Function}
\label{subsec:loss}
\vspace{-5pt}
Model parameters are optimized by minimizing a loss function computed as a modified version of the Euclidean distance between the SR image and the HR target. This modified loss function builds invariance to absolute brightness differences between ${I}^\text{SR}$ and ${I}^\text{HR}$.
Moreover, since ${I}^\text{SR}$ and ${I}^\text{HR}$ could be shifted, so the loss embeds a shift correction, resulting in:
\begin{align} \label{eq:loss}
L=\min_{u,v\in[0,2d]}\Vert{I_{u,v}^\text{HR}-({I}_\text{crop}^\text{SR}+b)}\Vert^2,
\end{align}
where ${I}_\text{crop}^\text{SR}$ is the cropped version of $I^\text{SR}$, $I_{u,v}^\text{HR}$ a shifted version of $I^\text{HR}$ and $b$ represents the brightness correction term:
$$b=\frac{1}{(rW-d)(rH-d)}\sum_{x,y} \left( I_{u,v}^\text{HR}-{I}_\text{crop}^\text{SR} \right).$$

\begin{table*}[ht]
\centering
\caption{Average mPSNR (dB) and SSIM}
\setlength\tabcolsep{4.8pt} 
\label{table:baselines}
\begin{tabular}{lccccccc}
& Bicubic+Mean & IBP \cite{IRANI1991231} & BTV \cite{1331445} & SISR+Mean & DUF \cite{Jo_2018_CVPR} & DeepSUM \cite{deepsum} & \textbf{DeepSUM++} \\ \hline
NIR & 45.69/0.97782 & 45.96/0.97960 & 45.93/0.97942 & 46.41/0.98166 & 47.06/0.98417 & 47.84/0.98578 & \textbf{47.93/0.98620} \\ \hline
RED & 47.91/0.98507 & 48.21/0.98648 & 48.12/0.98606 & 48.71/0.98787 & 49.36/0.98948 & 50.00/0.99075 & \textbf{50.08/0.99118} \\ \hline
\end{tabular}
\end{table*}

\begin{figure*}[t]
  \centering
    \begin{minipage}[b]{\textwidth}
        \begin{minipage}[c]{0.16\textwidth}
        \includegraphics[width=\textwidth]{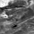}
        \end{minipage}
        \hfill
        \begin{minipage}[c]{0.16\textwidth}
        \includegraphics[width=\textwidth]{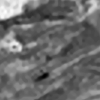}
        \end{minipage}
        \hfill
        \begin{minipage}[c]{0.16\textwidth}
        \includegraphics[width=\textwidth]{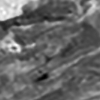}
        \end{minipage}
        \hfill
        \begin{minipage}[c]{0.16\textwidth}
        \includegraphics[width=\textwidth]{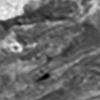}
        \end{minipage}
        \hfill
        \begin{minipage}[c]{0.16\textwidth}
        \includegraphics[width=\textwidth]{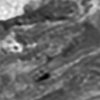}
        \end{minipage}
        \hfill
        \begin{minipage}[c]{0.16\textwidth}
        \includegraphics[width=\textwidth]{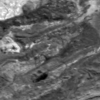}
        \end{minipage}
    \end{minipage}\\
    \vspace{-0.25cm}
    \caption{NIR band images (imgset1144). Left to right: one among the LR images, BTV(47.37 dB / 0.98284), DUF (48.02 dB / 0.98620), DeepSUM (48.74 dB / 0.98844), DeepSUM++ (49.46 dB / 0.98943), HR image.}
  \label{fig:zoom_images}
  \vspace{-0.5cm}
\end{figure*}

\vspace*{-8pt}
\section{RESULTS}
\label{sec:experiment}
\vspace*{-8pt}

In this section, we validate DeepSUM++ by comparing its performance with that of the following methods: averaged bicubic interpolated and registered images (Bicubic+Mean), CNN-based SISR method shared across multiple images followed by registration and averaging (SISR+Mean), IBP \cite{IRANI1991231}, BTV \cite{1331445}, dynamic upsampling filters (DUF) network \cite{Jo_2018_CVPR} and DeepSUM \cite{deepsum}. For all these methods, we followed the same procedure for the data preparation: bicubic interpolation and registration by phase correlation algorithm, except for DUF and DeepSUM that compute their own registration.

\vspace{-8pt}
\subsection{Experimental setting and training process}
\vspace{-5pt}
In the following experiments, we performed the same pre-processing and data preparation steps as in \cite{deepsum} for both NIR and RED band images. The images used to construct the two datasets have been captured by the PROBA-V satellite and are part the ESA challenge dataset \cite{web:kelvins}. %PROBA-V is an Earth observation satellite designed to map land cover and vegetation growth across the entire globe. 
The size of the collected images is $128 \times 128$ and $384 \times 384$ for the LR and HR data respectively. 

Before training DeepSUM++ as a whole, SISRNet-NL is pretrained by setting up a pure SISR problem (single input image) where an additional projection layer is added at the end, in order to turn the high-dimensional feature space into a single-channel image. SISRNet-NL with the final projection layer is trained with the same loss function in Eq. \eqref{eq:loss}.
DeepSUM++ is trained for 300 epochs with a batch size of 4, with separate models for NIR and RED. SISRNet-NL is initialized from the pretraining while FusionNet and RegNet weights are initialized from the model in \cite{deepsum}. 

We train DeepSUM++ using the minimum number of images available for each scene, i.e., $N=9$ images.

The exact number of network layers is shown in Fig. \ref{fig:Architecture} and the number of features is 64 everywhere except for the RegNet's first layer, which has 128 filters. Three graph convolutional layers are used in SISRNet-NL. 

\vspace{-8pt}
\subsection{Quantitative and qualitative results}
\vspace{-5pt}

The evaluation metric that we consider is a modified version of the PSNR (mPSNR), from which we derived the loss function described in Sec. \ref{subsec:loss}:
$$\text{mPSNR}=\max_{u,v\in[0,6]}20\log\frac{2^{16}-1}{\parallel{I_{u,v}^\text{HR}-({I}_\text{crop}^\text{SR}+b)}\parallel^2}$$
%The mPSNR computation is meant only for pixels that are not concealed both in the target HR image and in the reconstructed image. 
Similarly to the loss function, this metric has been devised to cope with the high sensitivity of the PSNR to biases in brightness and with the relative translation that the reconstructed image might have with respect to the target HR image.

%We remark that this metric was also used to evaluate submissions to the ESA challenge, where the score was computed as a ratio between the mPSNR of the submission and that of the baseline approach, average over all the held-out test set.

The validation has been performed over the same NIR and RED test sets used in \cite{deepsum}, using the sliding window procedure to use 13 images per scene in the testing process.
Table \ref{table:baselines} reports the results of the comparison. It can be noticed that the proposed method outperforms all the other methods. In particular, DeepSUM++ outperforms DeepSUM by 0.09 dB for NIR and 0.08 dB for RED band.
To ensure a fair comparison we have retrained DeepSUM following the same procedure used for DeepSUM++ for the same number of iterations.

These quantitative results are accompanied by a qualitative comparison in Fig. \ref{fig:zoom_images}. It can be noticed that our proposed method produces visually more detailed images, recovering finer texture and sharper edges.

\vspace*{-8pt}
\section{Conclusion}
\label{sec:conclusion}
\vspace*{-8pt}

In this paper, we demonstrated that non local information can be successfully exploited by neural networks to enhance the reconstruction quality of multitemporal remote sensing images in a MISR problem.

% References should be produced using the bibtex program from suitable
% BiBTeX files (here: strings, refs, manuals). The IEEEbib.bst bibliography
% style file from IEEE produces unsorted bibliography list.
% -------------------------------------------------------------------------

\end{document}